\documentclass[aps,prd,twocolumn,showpacs,superscriptaddress,groupedaddress,nofootinbib]{revtex4-1}  

\RequirePackage{ae}
\RequirePackage{bm} 
\RequirePackage{color}
\RequirePackage{amssymb}
\RequirePackage{amsmath}
\RequirePackage{amsfonts}
\RequirePackage{mathtools}
\RequirePackage{graphicx}
\RequirePackage{slashed}
\RequirePackage{wasysym}
\RequirePackage{setspace}
\RequirePackage{subfigure}
\RequirePackage{ulem}
\RequirePackage{placeins}

\RequirePackage{tabularx}
\RequirePackage{algpseudocode}
\RequirePackage{nicefrac}
\RequirePackage{bbm}
\RequirePackage{calligra}
\RequirePackage{microtype}
\RequirePackage{flafter}

\RequirePackage{tabularx}
\RequirePackage{ctable}

\newcommand{\dns}[1]{} 

\newcommand{\br}[1]{\left( #1 \right)}
\DeclareMathAlphabet{\mathcalligra}{T1}{calligra}{m}{n}

\newlength{\figlen}
\newlength{\figlenB}
\newlength{\figlenC}
\newlength{\figlenD}
\newlength{\tablen}
\newcount\Bool	
\def\eq$#1${\begin{multline}#1\end{multline}}

\newcommand{\txt}[2]{
  \ifnum\Bool=1 {#1} 
  \else	{#2}
  \fi
}

\begin{document} \sloppy

\title{Spin-states in multiphoton pair production for circularly polarized light}

\author{Christian Kohlf\"urst}
\email{C.kohlfuerst@gsi.de}
\affiliation{Theoretisch-Physikalisches Institut, Abbe Center of Photonics,
Friedrich-Schiller-Universit\"at Jena, Max-Wien-Platz 1, D-07743 Jena, Germany \\ 
Helmholtz-Institut Jena, Fr\"obelstieg 3, D-07743 Jena, Germany}


\begin{abstract}

Scalar and fermionic particle pair production in rotating electric fields is investigated in the nonperturbative
multiphoton regime. Angular momentum distribution functions in above-threshold pair production processes are calculated 
numerically within quantum kinetic theory and discussed
on the basis of a photon absorption model. The particle spectra can be understood if the spin states of the
particle-antiparticle pair are taken into account. 

 
\end{abstract}

\maketitle

\section{Introduction}

Photoelectron angular distributions (PAD) are well known in chemistry and atom physics, where the focus is on 
studying ionization spectra and understanding the inner structure of molecules and atoms \cite{Reid}. Viewing
multiphoton pair production as the highly relativistic analogue of the ionization of hydrogen, we
can expand the concept of angular momentum transfer via photon absorption to the strong-field QED regime.

In theory, particle production is similar to atomic ionization \cite{Brezin}, e.g., above-threshold effects in the multiphoton regime \cite{Delone:1994}.
Specifically, the particle under consideration absorbs more photons than necessary in order to transit to a continuous state, which manifests in a series of peaks in the momentum spectrum.
In both scenarios, the peak positions are determined by the number of absorbed photons and the field-dependent threshold \cite{Kohlfurst:2013ura,Lv:2018qxy,Otto:2014ssa,Akal:2014eua}.
A partial wave analysis, however, has been performed only in atomic physics, where it has been proven its importance towards understanding molecular
structures and the ionization process in general in various ways, 
see Refs. \cite{Reid,IonizeH} for reviews and further information on this topic as well as Ref. \cite{PhysRevA.87.033413} for a recent experimental verification.  

To be more specific, the angular dependence of the electron distribution carries information regarding the alignment of molecules \cite{PhysRevLett.86.1187}
as well as the impact of the core's Coulomb force on ionized electrons \cite{PhysRevLett.105.253002}.
Furthermore, with the aid of computer models the interference patterns in the electrons' momentum spectra could be examined thoroughly  
ultimately revealing details about the dynamics of photoionization as well as intramolecular dynamics \cite{Huismans61}.

With regards to strong-field matter creation, the focus has been on understanding the different mechanism to create particles in the first
place, see Refs. \cite{Sauter:1931zz}, \cite{Breit:1934zz}, \cite{Dinu:2017uoj} and \cite{Schutzhold:2008pz}. Furthermore, in contrast to atomic ionization there has been only one successful
experiment carried out so far \cite{Burke:1997ew}. Nevertheless, as laser development progresses the strong-field regime becomes more accessible, thus enabling studies on
particle creation processes on a regular basis, see Refs. \cite{Experiments, Ribeyre:2015fta} for detailed information on planned projects.

In this paper, we have investigated multiphoton pair production with focus on the angular dependence of the particle distribution. 
The distribution functions have been obtained using a phase-space formalism \cite{Vasak:1987umA,Heinz}, c.f. quantum kinetic theory (QKT) \cite{Smolyansky:1997fc}.
One big advantage of QKT is, that it is easy to obtain results for
fermionic as well as bosonic pair production, since both are described by a set of time-dependent ordinary differential equations. Additionally, the
framework allows to freely choose the polarization of the laser beams \cite{Li:2015cea,Rotating,Blinne:2015zpa}.

The particle momentum spectra are then analyzed on the basis of a photon absorption model taking into account energy and angular momentum conservation.
In this article we only consider simple, time-dependent, rotating electric fields \cite{Rotating,Blinne:2015zpa,VillalbaChavez:2012bb}. 
Such configurations provide the perfect setting for testing our absorption model. 
In particular, the orientation of the photon spin is fixed making selection rules much easier to apply \cite{Born}. 
In the following, we show the power and capabilities of the model by demonstrating how to identify the imprint the different spin states 
leave on the electron-positron angular distribution.
 
%
%
%
%
%

Throughout this paper we use natural units $\hbar=c=1$ and measure all dimensionful quantities in terms
of the mass of the particles $m$.

\section{Electric field pulse}

In order to study pair production for circularly polarized light, we have chosen the following model for the vector potential
\begin{equation}
 \mathbf{A} \left( t \right) =  -\frac{\varepsilon E_{cr}}{2\omega} \exp{ \left(-\frac{t^2}{\tau^2} \right) } 
 \begin{pmatrix}
  \sin \left( \omega t \right) \\
  \cos \left( \omega t \right) \\
  0
 \end{pmatrix}         ,
 \label{eq:A}
\end{equation}
with the critical field strength $E_{cr}=m^2/e$, the peak field strength $\varepsilon$, the pulse length $\tau$ and the photon energy $\omega$. 
The electric field is derived from this expression.

We have chosen the model \eqref{eq:A} to mimic a standing wave formed by two counter-propagating laser beams with propagation
direction $\pm \mathbf{\hat z}$. To form a standing wave pattern as given by eq. \eqref{eq:A} one beam has
to be left-handed and the other must be right-handed. As photons are spin-1 bosons we assume without loss of generality the
helicity to be $+1$ \cite{Born}. Hence, every photon not only transfers an energy quant of $\omega$ to the particle-antiparticle pair, but also
increases its total angular momentum by one. 

\section{Quantum Kinetic Theory}

All numerical results are based upon a phase-space approach belonging to a particular class
of quantum kinetic theories. As we perform calculations for spatially homogeneous fields the governing equations of motion
take on a numerically favorable form. As a result, we are able to obtain very accurate numerical solutions allowing us to compare
the computed data with predictions from the model.

In the Weyl gauge, the quantum kinetic equations for spin-$1/2$ particles can be written as \cite{Rotating}
\begin{alignat}{5}
    & \partial_{t} \mathbbm{s}^v    && && -2 \mathbf{p} \br{t} \cdot \mathbbm{t_1}^v && &&= 
      - 2 \frac{ \ Q \br{t} }{ \omega \br{t}}, \\    
    & \partial_{t} \boldsymbol{\mathbbm{v}}^v && && +2\mathbf{p} \br{t} \times \boldsymbol{\mathbbm{a}}^v && +2\mathbbm{t_1}^v &&= 
       2 \frac{ e \mathbf{E} \br{t} - Q \br{t} \mathbf{p} \br{t} }{\omega \br{t}}, \\    
    & \partial_{t} \boldsymbol{\mathbbm{a}}^v && && +2 \mathbf{p} \br{t} \times \boldsymbol{\mathbbm{v}}^v && &&= \mathbf{0},  \\
    & \partial_{t} \boldsymbol{\mathbbm{t_1}}^v && && +2 \mathbf{p} \br{t} \ \mathbbm{s}^v && -2 \boldsymbol{\mathbbm{v}}^v &&= \mathbf{0}  
\end{alignat}
with initial conditions
\begin{align}
  \mathbbm{s}_i^v = 0,\quad \boldsymbol{\mathbbm{v}}^v_{i} = \boldsymbol{\mathbbm{a}}^v_{i} = \boldsymbol{\mathbbm{t_1}}^v_{i} = \mathbf{0}.
\end{align}  
Following Refs. \cite{Vasak:1987umA, Heinz} we can associate the individual terms in the differential equation as mass density $\mathbbm{s}^v$, 
current density $\boldsymbol{\mathbbm{v}}^v$, spin density $\boldsymbol{\mathbbm{a}}^v$ and magnetic moment density $\boldsymbol{\mathbbm{t_1}}^v$.
We have used abbreviations for the one-particle energy $\omega \br{t} = \sqrt{1+ \mathbf{p} \br{t}^2}$, 
the kinetic momentum $\mathbf{p} \br{t} = \mathbf{q} -e \mathbf{A} \br{t} $ and the auxiliary variable 
$Q \br{t} = \dfrac{ e \mathbf{E} \br{t} \cdot \mathbf{p} \br{t} }{ \omega \br{t}^2}$. 

The equations of motion for scalar particles, on the other hand, are given by \cite{Heinz}
\begin{alignat}{5}
    & \partial_{t} \mathbbm{f}^v    && &&  -\mathbf{p} \br{t}^2 \left( \mathbbm{g}^v + \mathbbm{h}^v \right) && -2 \mathbbm{g}^v &&= 
      2 \frac{\mathbf{p} \br{t}^2}{\omega \br{t}}, \label{eq:sQED1} \\   
    & \partial_{t} \mathbbm{g}^v    && &&  +\mathbf{p} \br{t}^2 \mathbbm{f}^v && +2 \mathbbm{f}^v &&= 
      -\frac{1}{2} \frac{\left(1+ \omega \br{t}^2 \right) Q \br{t} }{\omega \br{t}}, \\   
    & \partial_{t} \mathbbm{h}^v    && &&  -\mathbf{p} \br{t}^2 \mathbbm{f}^v && &&= 
      -\frac{1}{2} \frac{\mathbf{p} \br{t}^2 Q \br{t} }{\omega \br{t}}.  \label{eq:sQED3}       
\end{alignat}
Again, the initial conditions have been incorporated into the equations of motion, thus
\begin{align}
  \mathbbm{f}_i^v = \mathbbm{g}_i^v = \mathbbm{h}_i^v = 0.
\end{align}  
From Refs. \cite{Heinz} we know to identify $\boldsymbol{\mathbbm{j}}^v = \mathbf{p} \br{t} \left( \mathbbm{g}^v + \mathbbm{h}^v \right)$ as current density and 
$\mathbbm{h}^v$ as mass density. The term $\mathbbm{f}^v$ does not seem to have a direct physical interpretation.


Throughout this article we will discuss the results on the basis of the spin-dependent particle number density $f_s$, which is evaluated at asymptotic times. 
Hereby, the electron-positron distribution function is given by 
\begin{align}
f_{1/2} \left( p_x, p_z \right) = \frac{\mathbbm{s}^v + \mathbf{p} \cdot \boldsymbol{\mathbbm{v}}^v}{\omega} \bigg \vert_{t \to \infty} \label{equ:n}
\end{align}
and the particle number density for spinless particles is given by
\begin{align}
f_0 \left( p_x, p_z \right) = \frac{2 \mathbbm{h}^v + \mathbf{p} \cdot \boldsymbol{\mathbbm{j}}^v }{\omega} \bigg \vert_{t \to \infty}. \label{equ:f}
\end{align}
In the parameter region of interest ($\varepsilon \ll 1,~ \omega \approx \mathcal{O}(m),~ \omega \tau \gg 1$) the particle momentum spectrum at asymptotic times (vanishing vector potential) is axially symmetric in $(p_x,p_y)$. 
For the sake of simplicity, we therefore have $p_y=0$ and refer
to $p_x$ as in-plane or parallel momentum. In this regard, $p_z$ defines the perpendicular momentum.

\section{Photoparticle angular distribution}   

In order to enable multiphoton particle production the total photon energy $n \omega$ has to exceed not only the particles' rest mass but also
their oscillatory energy in the background field $2 m_\ast$. Here, $m_\ast$ describes the particle's effective mass
in a homogeneous, oscillating background field
\begin{equation}
 m_\ast = m \sqrt{1+ \xi^2 } \approx m \sqrt{1+ \frac{\varepsilon^2 m^2}{4 \omega^2} },
\end{equation}
where $\xi=\frac{e}{m} \sqrt{ -\langle A^\mu A_\mu \rangle }$.
Moreover, energy conservation laws dictate the particle's kinetic energy and as a consequence its momentum reads \cite{Kohlfurst:2013ura}
\begin{equation}
 \mathbf{p}^2_\ast = \left(\frac{n \omega}{2} \right)^2 - m_\ast^2. \label{equ:En1}
\end{equation}
From eq. \eqref{equ:En1} we can already deduce that particle distributions form shells in phase space, which can be
classified by the number of photons absorbed \cite{Otto:2014ssa}. 
The effective mass model, however, cannot explain why each shell falls off differently.

Within the absorption model, given a specific shell $n$, the particle's angular distribution results from squaring the photoelectron wavefunction, which, in turn, 
can be generically expanded in terms of spherical harmonics $Y_l^m$ 
\begin{multline}
 I \left(\vartheta, \varphi \right) \propto \psi^{\ast} \psi \\ = \sum_{l=0}^{n} \sum_{l'=0}^{n} \sum_{m=-l}^l \sum_{m'=-l'}^{l'} b_{l'm'}^{\ast} b_{lm} 
  Y_{l'm'}^{\ast} Y_{lm}, \label{equ:I}
\end{multline} 
with the coefficients $b_{lm}$ and the angular momentum quantum numbers $l$ and $m$. Naturally, the particle's angular momentum is
limited by the number of photons absorbed. 
For the case under consideration, however, it is useful to perform the partial wave analysis for the particle distribution and not the photoelectron wavefunction \cite{Yang:1948nq}
\begin{equation}
 I \left(\vartheta, \varphi \right) \propto \sum_{L=0}^{2n} \sum_{M=-L}^L B_{LM} Y_{LM} \left(\vartheta, \varphi \right), \label{equ:I2}
\end{equation}
where $L$ and $M$ are based upon $l$, $l'$ and $m$, $m'$, respectively.

      \begin{figure}[t]
      \begin{center}
	\includegraphics[width=0.45\textwidth]{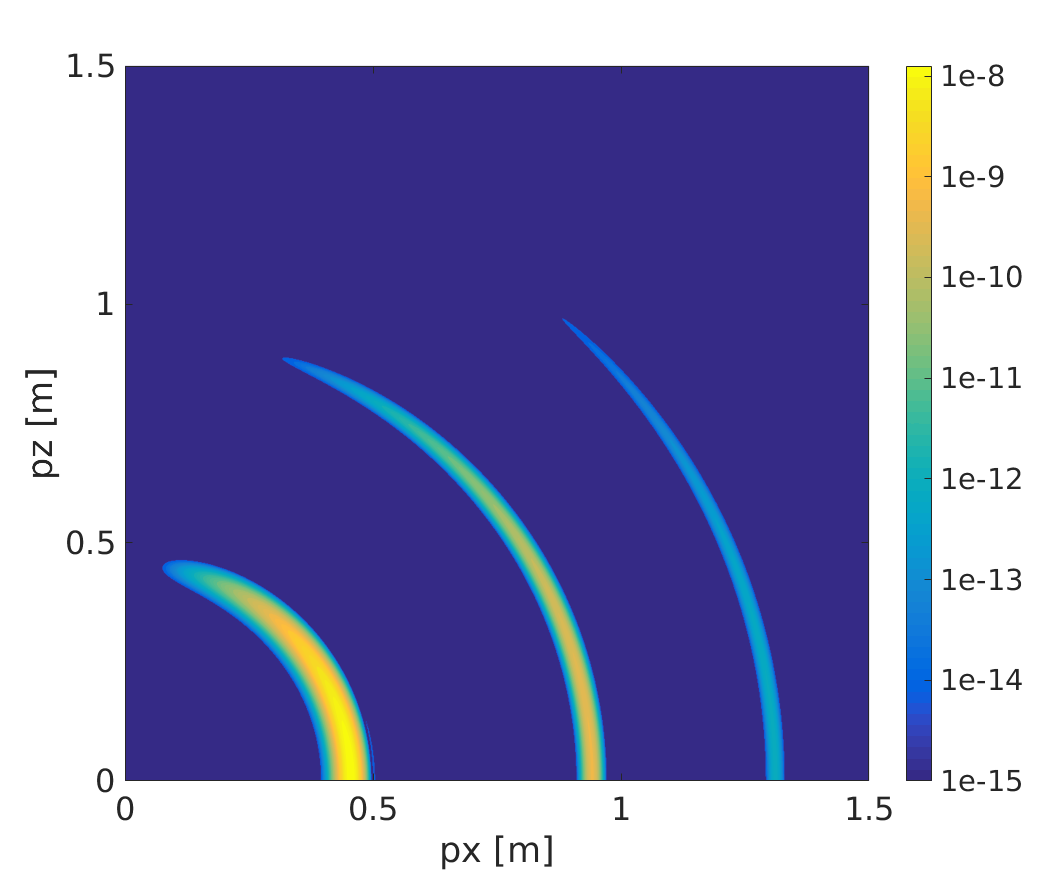} 	
      \end{center}
      \caption{Scalar particle distribution function $f_0$ as a function of the in-plane momentum $p_x$ and the perpendicular momentum $p_z$ for
      a field with peak strength $\varepsilon=0.075$, frequency $\omega=0.55m$ and pulse length $\tau=250m^{-1}$.
      The shells arise due to the absorption of $4$, $4+1$ and $4+2$ photons (above-threshold pair production).}
      \label{fig:fpxpz}
      \end{figure}   

As stated in the beginning, the model for the background field \eqref{eq:A} describes photons with helicity $+1$ only \cite{1967JETP...25.1135N}. 
Hence, in an $n$-photon process the total angular momentum before particle creation takes place is equal to the number of photons $n$ in the process. 
As angular momentum must be conserved, the total angular momentum of the produced particle pair has to be $n$, too. Hence, every photon absorbed increases the 
total angular momentum of the particles by one. However, as the pair's angular momentum is composed of an intrinsic spin component and an orbital component,
a fraction of the transfered photon angular momentum might be needed in order to create the particle's spin in the first place, c.f., section on 
electron-positron pair production. 
Additionally, all photons have the same helicity thus $m=l$ and consequently $M=L$ holds.

For circularly polarized light eq. \eqref{equ:I2} therefore takes on the remarkably simple form
\begin{equation}
 I_S \left(\theta \right) = M_S \sin^{2(n-S)} \left( \theta \right), \label{I_c}
\end{equation}
where $S$ describes the pair's intrinsic particle spin and $M_S$ states the spin-dependent creation rate. The angle
$\theta$ is defined as the angle between the field's propagation direction and the particle's ejection direction.


\section{Scalar pair production}

The shell structure as well as the lack of pronounced interference patterns can be observed in our results. In Fig. \ref{fig:fpxpz}, where we employed a background
field with peak strength $\varepsilon=0.075$, frequency $\omega=0.55m$ and pulse length $\tau=250m^{-1}$,
we recognize a perfectly regular pattern of three peaks at momenta $\mathbf{p}_4 = 0.4556m$, $\mathbf{p}_5 = 0.9419m$ and $\mathbf{p}_6 = 1.311m$. 
Here, the indices enumerate the shells with above-threshold peaks
caused by the absorption of $4$, $5$ and $6$ photons. The predictions obtained by the effective mass model yield $\mathbf{p}_{\ast,4}=0.4532m$, 
$\mathbf{p}_{\ast,5}=0.9413m$ and $\mathbf{p}_{\ast,4}=1.3107m$, which support our interpretation.      

      \begin{figure}[t]
      \begin{center}
	\includegraphics[width=0.49\textwidth]{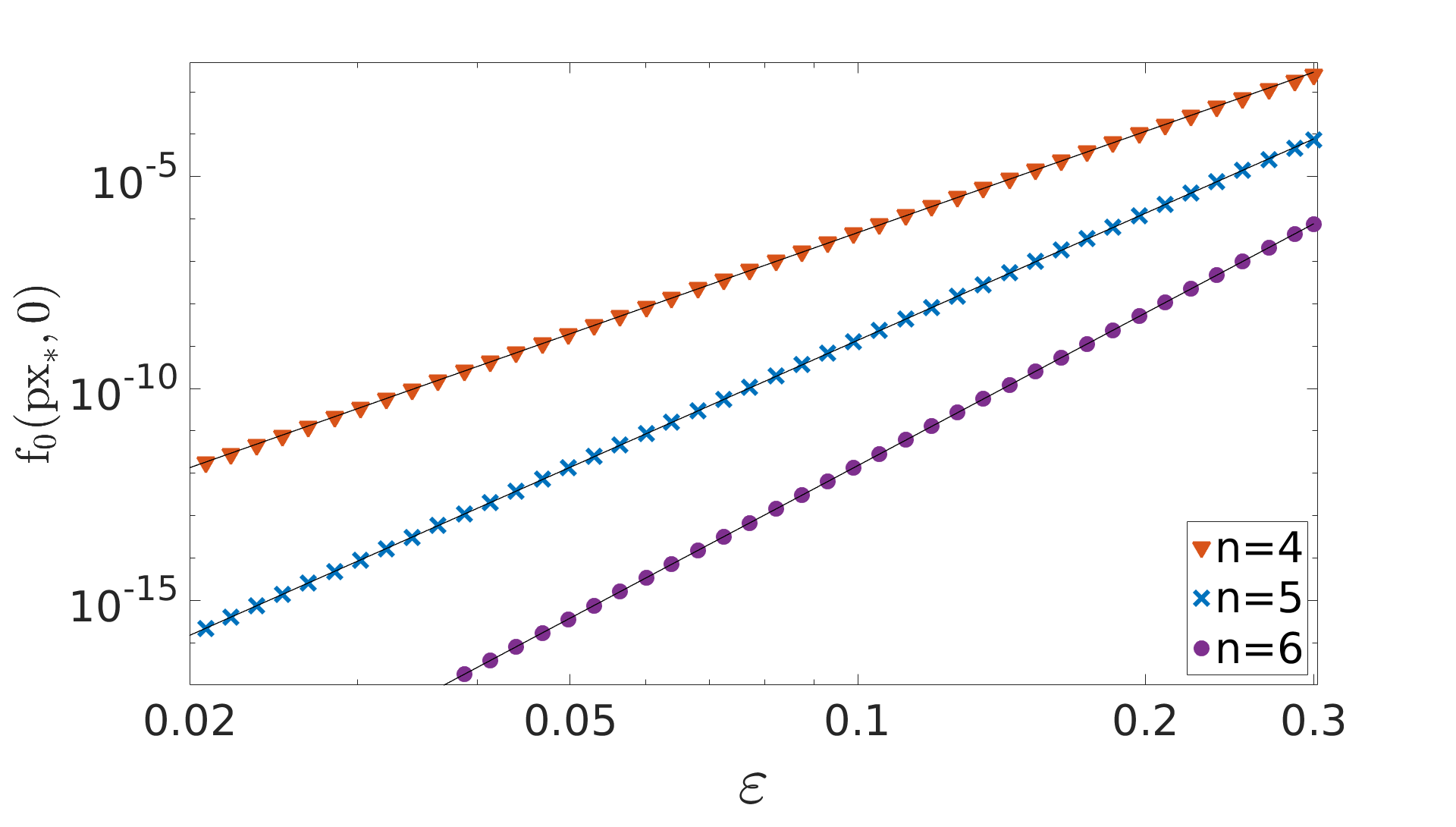} 	
      \end{center}
      \caption{Displaying peaks in the $n$-photon production probability as a function of the background field strength $\varepsilon$ for $\omega=0.666m$ and $\tau=250m^{-1}$.
      Despite a field-dependent threshold we can monitor the local maxima in the spectrum $f_0 (p_{x,\ast},0)$ and fit a model of the form $\varepsilon^{2n}$
      to the numerical data (black lines).}
      \label{fig:f_eps}
      \end{figure}  

As we also aim to obtain a quantitative understanding of the processes, we test for non-trivial scaling of the $n$-photon peaks. 
Hence, we have solved the governing equations of scalar QKT \eqref{eq:sQED1}-\eqref{eq:sQED3} for various values of the peak field strength $\varepsilon$ 
at the effective momenta $\mathbf{p}_\ast = (p_{x,\ast},0)$ for $4$-, $5$- and $6$-photon processes, c.f. Fig. \ref{fig:f_eps}. Monitoring each peak individually
has the advantage that we can ensure we are probing a fundamental observable; in Ref. \cite{Kohlfurst:2013ura} the total yield was discussed, which suffers 
from having multiple sources of contributions. We fit a simple model of the form of $\varepsilon^{x_n}$ to the data, because
the intensity of a laser beam coincides with the number of photons in the beam $I \sim \varepsilon^{2} \propto n$.
The exponents were calculated to be $x_4=7.94$, $x_5=9.96$ and $x_6=11.96$ with confidence intervals of $(7.92, 7.95),~ (9.96, 9.97)$ and $(11.96, 11.97)$ at $95\%$ confidence.
For comparison, our model predicts an increase of $\varepsilon^{2n}$, thus de facto excluding any non-trivial effects at low field strengths. 

\section{Electron-positron pair production}
              
Electron-positron momentum spectra generally look similar compared to their scalar counterparts.
The differences stem from the fact that there are two options for spin alignment in a pair of spin-$1/2$ particles; parallel and anti-parallel.
In the former, both spins point in the same direction thus their total spin is given by $S=1/2+1/2=1$; while in the latter the total spin adds up to $S=0$.
Hence, in an $n$-photon absorption process there are now two options for the 
particles orbital angular momentum. As a result, the model for fermionic pair production is given by the sum of the two contributions, c.f. eq. \eqref{I_c},
\begin{equation}
 I \left(\theta \right) = 
 \left| M_0 \sin^{2n} \left( \theta \right) + M_1 \sin^{2(n-1)} \left( \theta \right) \right|, \label{I_ferm}
\end{equation}
where absolute values are introduced to ensure nonnegativity, because $M_i$ can take on negative values. 
This is slightly different compared to the simpler case of scalar pair production, where no $S=1$ state exists.

      \begin{figure}[t]
      \begin{center}
	\includegraphics[width=0.49\textwidth]{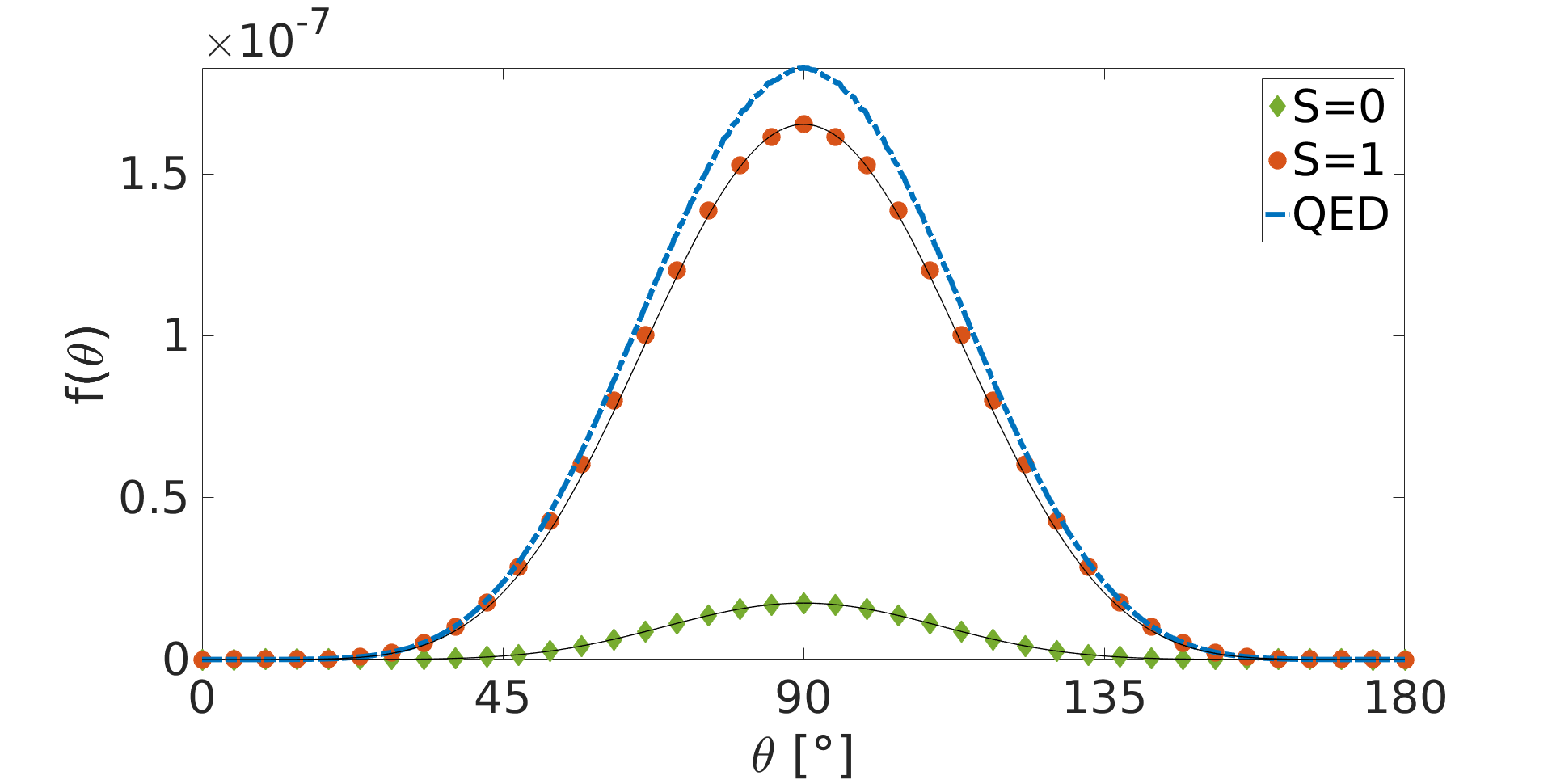} 	
      \end{center}
      \caption{Photoelectron angular momentum $f(\theta)$ in case of $4$-photon pair production decomposed in terms of anti-parallel ($S=0$) and parallel ($S=1$) spin contributions 
      for $\varepsilon=0.075$, $\omega=0.55m$ and $\tau=250m^{-1}$. The black lines correspond to fits of the form of $\sin^{2(4-S)} (\theta)$.
      Scalar pair production is given by the green curve (diamonds).      
      Electron-positron pair production (QED) is the sum of scalar and spin-1 contributions.}
      \label{fig:ftheta}
      \end{figure}     

One consequence of the existence of an excited spin state becomes apparent when investigating particles with vanishing momentum $\mathbf {p}$.
Every photon absorbed by the particle pair adds $+1$ to its total angular momentum. In case of scalar pair production, or if spins are aligned
anti-parallel ($S=0$), all angular momentum is carried by the quantum number $L$. As a non-zero orbital momentum requires a non-zero linear momentum,
pair production cannot happen in this case. The only option for particle creation at vanishing linear momentum $\mathbf{p}$ is a one-photon process, 
where the particle spins align parallel, thus absorbing the photon angular momentum and resulting in $L=0$ \cite{Marinov:1977gq}.

Interestingly, one-photon absorption is also the only way to obtain particle ejection in the propagation direction of the laser beams. If the energy
of one and only one photon exceeds the rest energy of a fermionic particle pair, we obtain a closed shell in momentum space. This is because only 
for the one-photon process an electron-positron pair in an $S=1$ spin state can absorb all the photon angular momentum and have a 
vanishing orbital angular momentum.

Generally, there are contributions from both spin-states to the electron-positron momentum spectrum. 
In this regard, it is interesting to discuss the particle spectrum as a function of the angle $\theta$, where $\theta=90^\circ$ corresponds to $(p_x,0)$, 
see Fig. \ref{fig:ftheta}.

Solving scalar QKT for a $4$-photon event ($\varepsilon=0.075$ and $\omega=0.55m$) we obtain a particle distribution ($\mathbf{p}_\ast=0.456m$) which is
perfectly described by our model; here $\overline M_0 = 0.9987 \ (0.998, 0.999)$ with $95\%$ confidence interval and $R^2=0.999997$ using normalized quantities.
In the case of fermionic particles, we obtain a distribution function $f_{1/2} (\theta)$ with contributions from an $S=0$ as well as from an $S=1$ state.
Assuming that for a given field configuration the particle creation rates for the attainable spin states do not change (besides a trivial factor of $2$
due to combinatorics in $M_0$), we can subtract the zero-spin production probability from the total probability yielding the contribution from the $S=1$ state alone.

Applying our model to fit the data we find the numerical results being well described by a ${\rm sin}^{6}$ function, as 
we obtain $\overline M_1 = 1.01$ with confidence interval $(1.00, 1.02)$ at $95 \%$ certainty and $R^2=0.9994$.
It is remarkable to see the
contribution from the $S=1$ state being much higher than the contribution from the singlet state. In a way, this situation
resembles the status regarding orthohelium and parahelium with orthohelium ($S=1$) having a lower rest energy.

\section{Discussion}

Particle pair production is a complex process and observables are very sensitive to changes in the background fields.
Nevertheless, it seems as if we have found a way to understand certain features of multiphoton pair production intuitively.

The specific procedure in which we obtained the particle creation rates $M_i$ is, in its current form, only applicable to purely circularly polarized fields.
Although the generic model \eqref{equ:I} can be applied for arbitrary field configurations, the composition of the particle spectra in terms
of angular momentum contributions takes on the particularly simple form \eqref{I_c} only in case of rotating fields.
For linearly polarized light, for example, the particles final angular momentum becomes a sum over many terms, because the
initial photons could have been left- or right-handed.

Our model is also limited to the nonperturbative multiphoton regime in pair production for Keldysh parameter $\gamma_\omega = \omega / \left(m \varepsilon \right) > 1$.
Moreover, the studied momentum signatures only emerge for multi-cycle pulses $\omega \tau \gg 1$. In a few-cycle pulse, the individual photon energies 
are varying too much, thus the clear characteristic peaks cannot form in the final particle spectrum. 


Finally, charge-conjugation and parity invariance pose additional constraints on multiphoton pair production. 
For an even (odd) number of photons $n$, the (scalar) momentum distribution has to vanish for vanishing momentum $\mathbf{p}$ independent of the polarization of the incoming beams.
These constraints become especially important for linearly polarized fields, where, in principle, the angular momentum transfer-picture allows for
pair production at $\mathbf{p}=0$ for higher photon numbers. Within the context of this work, the model coincides with all selection rules.

\section{Summary}

We have demonstrated that, in multiphoton pair production, the distribution functions for scalar and spin one-half particles in rotating electric fields
follow a specific, intuitive pattern. On the basis of accurate numerical solutions of quantum kinetic theory it is straightforward to obtain the spin-dependent particle
creation amplitudes. Although the procedure presented in this article is based on circularly polarized waves, the underlaying model 
can be readily extended to arbitrary field configurations.

\section{Acknowledgments}

We thank Andr\'e Sternbeck, Holger Gies, Alexander Blinne and Greger Torgrimsson for interesting discussions.
We thank Holger Gies and Se\'{a}n Gray for a critical reading of the manuscript.
The work of C.K.~is funded by the Helmholtz Association through the
Helmholtz Postdoc Programme (PD-316). We acknowledge support by the
BMBF under grant No. 05P15SJFAA (FAIR-APPA-SPARC).


\bibliographystyle{unsrt}

\end{document}